\documentstyle[preprint,aps]{revtex}
\begin{document}
\draft

\title{Scaling of level statistics at the disorder induced
metal-insulator transition\thanks{to appear
in Phys. Rev. B {\bf 51}, 17239 (1995)}
}

\author{I. Kh. Zharekeshev\cite{perIZ} and B. Kramer\\ }

\address{ 1. Institut f\"ur Theoretische Physik, Universit\"at Hamburg,
Jungiusstrasse 9, D--20355 Hamburg, Germany }

\date{\today}
\maketitle

\begin{abstract}
The distribution of energy level separations for lattices of sizes
up to 28$\times$28$\times$28 sites
is numerically calculated for the Anderson model.
The results show one-parameter scaling.
The size-independent universality of the critical level spacing
distribution allows to detect with high precision
the critical disorder $W_{c}=16.35$. The scaling properties
yield the critical exponent, $\nu =1.45 \pm 0.08$,
and the disorder dependence of the correlation length.
\end{abstract}

\pacs{PACS numbers: 71.30.+h, 05.45.+b, 72.15.Rn}

The statistical properties of the energy spectra of disordered quantum systems
in the metallic regime can be described by
the random-matrix theory initiated by
Wigner and Dyson~\cite{Wigner55,Dyson62,RMT}.
This was shown by analytical approaches
such as the supersymmetric nonlinear $\sigma$-model formalism~\cite{Efetov83}
and the impurity diagram technique~\cite{Altshuler86}.
A characteristic property in this regime
is the repulsion between adjacent energy levels which is caused
by a pronounced overlap of delocalized one-electron states.

By increasing the fluctuations
of the random potential in three-dimensional (3D) systems
a delocalization-localization transition is induced,
the Anderson metal-insulator transition (MIT)~\cite{Lee85}.
In the insulating phase,
when the magnitude of the fluctuations
of the potential is much larger than the critical value,
the states of the system are localized.
The corresponding energy levels
are not correlated due to the vanishing
of their spatial overlap.
Therefore, spectral fluctuations
are governed by the statistics of independent random variables.
A central question in discussion
of the level statistics in  disordered systems
is how the spectral fluctuations change their behavior
when extended states are transformed into
localized ones when the disorder is increased.

According to the one-parameter scaling
theory~\cite{AARL79} the conductance of the system shows
critical behavior at the MIT.
Since the conductance is related to the spectral
two-level correlation function
one can expect that the level statistics also experiences the
crossover between the metallic and the insulating
regimes~\cite{Altshuler88}.
Correlations of the energy levels are contained in the distribution
of the distances between successive energy levels, $P(s)$.
In the metallic region $P(s)$
can be well described by the Wigner surmise
\begin{equation}
P_{W}(s)=\frac{\pi}{2}\,s\,\exp \left({-\frac{\pi}{4} s^2}\right),
\label{Wigner}
\end{equation}
proposed primarily for the Gaussian orthogonal ensemble of random
matrices whose elements fluctuate normally
around zero with the equal variances~\cite{Wigner55,RMT}.
Here the distances between neighboring energy levels $s$ are measured
in units of the mean level spacing $\Delta$.
In the localized regime the spacings are distributed according
to a Poisson law
\begin{equation}
P_{P}(s)=\exp (- s).
\label{Poisson}
\end{equation}
The crossover from the metallic to the insulating regime is accompanied by a
change in the level spacing distribution from correlated to completely
uncorrelated behavior. One can, therefore, suspect that the properties of
$P(s)$  for finite systems, when investigated as a function of the size of
the system, would allow for the detection of the MIT and even the critical
properties of the transition.
As a consequence, the crossover of $P(s)$
between the Wigner~(\ref{Wigner}) and the Poisson~(\ref{Poisson}) statistics
was extensively studied both analytically~\cite{Imry87,Aronov94}
and numerically
\cite{Altshuler88,Shklovskii93,Evangelou92,Hofstetter93,Hofstetter94}.
By using of the finite-size scaling argument for $P(s)$
a new criterion for the location of the metal-insulator
transition was proposed by Shklovskii {\it et al}~\cite{Shklovskii93}.
This argument was cross-checked by numerically determining level statistics,
and investigating the scaling properties for large $s$.
One advantage of this method is that one needs only the energy spectrum
without calculating eigenfunctions or the conductivity.

In this paper we focus again on the critical behavior of the level
statistics near the disorder-induced MIT. We study
the finite-size scaling  properties
$P(s)$ in the whole range of spacings.
For the calculation of the energy levels,
numerical diagonalization of the Anderson model
was performed by using a modified Lanczos algorithm
for systems of sizes up to 28$\times$28$\times$28 lattice sites.
We determine very precisely the critical value of the disorder
$W_{c}$ and the correlation length exponent $\nu $.
By considering the integrated distribution of spacings in various regions
of $P(s)$, we find $W_{c}=16.35$ and $\nu=1.45\pm 0.08$ in good
agreement with similar earlier calculations~\cite{Hofstetter94}
and previous findings based on the transfer-matrix (TM) method for the
modulus of the Green's function~\cite{Kramer83,KramerM94}.
We provide an independent determination
of the correlation length as a function of the disorder, $\xi (W)$.
Our results demonstrate unambiguously
and strikingly, that the integrated level spacing distribution
plays not only the role of a scaling variable at the Anderson transition,
but can also be used to define qualitatively and accurately the critical
properties.

The Anderson model for a disordered system
is described by the Hamiltonian
$H=\sum_{n}\varepsilon_{n}^{} c_{n}^{\dag} c_{n}^{} +
          \sum_{n\neq m} (c_{n}^{\dag} c_{m}^{} + c_{n}^{} c_{m}^{\dag})$,
where
$c_{n}^{\dag}$ and $c_{n}^{}$ are the creation and annihilation operators of
an electron at a site $n$ in a lattice;
$m$ denotes the number $Z$ of the nearest neighbors of the site $n$
($Z=6$ in a 3D simple cubic lattice).
The site energies $\varepsilon_{n}$
measured in units of the overlap integral between adjacent sites
correspond to the random potential. They are
uniformly distributed in the interval [$-W/2,W/2$],
where $W$ specifies the degree of the disorder of the system.
It was found earlier by the TM-method~\cite{Kramer83} that
the critical disorder of the MIT which occurs in the middle of the band $E=0$
corresponds to $W_{c}\approx 16.5$.
Thus, for a finite system when $W$ decreases from $W_{c}$,
the level statistics should approach (\ref{Wigner}),
and when $W\gg W_{c}$ it should obey~(\ref{Poisson})~\cite{Altshuler88}.
In order to find the electron spectrum in the critical region
for $W$ close to $W_{c}$
we diagonalized numerically the real symmetric
matrices of the Anderson Hamiltonian
for simple cubic lattices of the size $L^3$ with periodic boundary conditions.
One has to consider system sizes much larger than the lattice constant.
This is necessary in order to get rid of the
microscopic peculiarities of the discrete lattice model,
and to be able to compare numerical with analytical results
obtained in the thermodynamic limit.
Despite $H$ is represented  by a very sparse matrix
the computation of the energy levels of large
systems is highly nontrivial due to the huge matrix size which grows as $L^6$.

We have developed a considerably improved Lanczos algorithm for
the eigenvalue problem of very sparse and big matrices.
One of the improvements is based on the hierarchical structure of $H$
which consists of three levels of complexity
depending on the dimensionality of the lattice.
Another feature
is that all non-zero off-diagonal matrix elements are equal to unity.
This enabled us to decrease substantially
the number of multiplications during the recursion~\cite{Cullum85}.
In addition, the absence of degeneracy of the eigenvalues
leads to a faster convergence of the eigenvectors.
We exploited the effect of level repulsion in the range of
$s\lesssim 0.1$ for diminishing the
number of steps of the Sturm sequence in the bisection method
that is used for searching the eigenvalues
of the tridiagonal matrix produced after the recursion.
In this case one can govern the precision conditions for resolving
two very close `real' eigenvalues provided that the absolute
accuracy for recognizing `spurious' eigenvalues is kept as high
as of the order of $10^{-12}$.
All of these optimizations enabled us to create
a computer program which consumes much less
time than a conventional implementation of the Lanczos method~\cite{Cullum85}.
As a result, we succeeded in diagonalizing matrices of sizes up to
$21952\times 21952$.
We applied the algorithm to cubes of linear sizes
$L = 6, 8, 12, 16, 20, 24, 28$ at various degrees of the disorder
close to $W_{c}$.

One consequence of the fact that the distribution of
$\varepsilon_{n}$ is box-like is that the correlation length $\xi$
depends only very weakly on the energy near the band center~\cite{KramerM94}.
This follows from the phase diagram of localization which shows that
$W_{c}(E)$ is almost constant when the mobility edge
varies  from $E=-6$ to $6$.
This allows us to use a wide central part of the band containing
half of all of the eigenvalues of $H$,
without mixing localized and delocalized states.
One can check that the level
statistics $P(s)$ for narrower energy intervals do not change considerably.
The calculations were carried out for
ensembles of different random configurations.
For a given pair of \{$L,W$\} the number of samples
was chosen such that a total of approximately $10^5$ eigenvalues were
taken into account.
Since the energy interval is so wide, the density of states
$\rho = (\Delta L^3)^{-1}$ depends on energy, though only slightly,
one has to unfold the spectrum.
The unfolding was performed by using parabolic spline fits for $\rho(E) $.

As an example, Fig.~\ref{fig1}\, displays the histograms of $P(s)$
for $L=6$ and $28$ at  various disorder $W$ near the MIT.
By increasing $W$ the spacing distributions for both $L$ vary
continuously from $P_{W}(s)$ to $P_{P}(s)$,
particularly in the range of $s\gtrsim 2$.
Results for other, intermediate, sizes confirm this smooth crossover of $P(s)$.
However, the behavior of the crossover depends substantially on
the size of the cube.
One can observe in Fig.~\ref{fig1} that $P(s)$ changes
from~(\ref{Wigner}) to~(\ref{Poisson}) faster for $L=28$ than for $L=6$.
As the size of the system
increases the function $P(s)$ approaches the above two
limiting regimes within a narrower interval of $W$ around $W_{c}$.
The size-dependence of $P(s)$ is observed on both sides of the MIT.
Close to $W=16.5$, the spacing distribution has almost the same form
for all considered $L$, from 6 up to 28.
The independence of $P(s)$ on $L$ at $W=W_{c}$ is
consistent with the prediction of a novel universal
level statistics at $L\to \infty $
that exists exactly at the transition~\cite{Shklovskii93}.
One can expect that in the thermodynamic limit
there are three asymptotic shapes,
namely, the Wigner surmise~(\ref{Wigner}) for $W < W_{c}$;
the Poisson law~(\ref{Poisson})
for $W>W_{c}$, and the critical distribution $P_{c}(s)$ when $W=W_{c}$.

Since $P(s)$ scales differently with $L$ for various regions of $s$,
it is more convenient to consider the integrated probability
$I(s)=\int_{s}^{\infty}\,P(s^{\prime})\,ds^{\prime}$,
which equals a fraction of those spacings which are larger than $s$.
Because all spacings are positive, $I(0)=1$.
In addition, $\int_{o}^{\infty}\,I(s)\,ds=1$ due to
normalization to the total number of spacings in a given energy interval.
Eqs.~(\ref{Wigner}) and~(\ref{Poisson}) give
$I_{W}(s)=\exp(-\pi s^2/4)$ and $I_{P}(s)=\exp(-s)$
for the two limiting regimes. For numerical purposes we introduce
$J(s) = (I(s)-I_{W}(s))/(I_{P}(s)-I_{W}(s))$.
It describes the relative deviation of $I(s)$ from $I_{W}(s)$.
In order to extract the dependence
only on $W$ and $L$ and to minimize numerical errors
we define a certain  value of the spacing $s_{o}$ for $J(s)$.
The choice of the average spacing, $s_{o}=1$,
is not favorable, because  $J(\Delta )$ exhibits a very small size effect.
One can see in  Fig.~\ref{fig1} that $P_{W}(s)$ and $P_{P}(s)$ cross at
$s_1=0.473$ and $s_2=2.002$.
Therefore we studied $\alpha(W,L) \equiv J(s_1)$
representative for the small-$s$ part of $P(s)$,
and  $\gamma(W,L) \equiv J(s_2)$ representative for its tail,
where $I_{W}(s)$ and $I_{P}(s)$ are significantly different.
The former reflects the ``strength'' of the  repulsion of
two consecutive levels when the separation between them is less than $s_1$,
while the latter is the exponentially small probability to find a
gap in the spectrum which is wider than two mean spacings $\Delta $.
For $L\rightarrow \infty$ both, $\alpha$ and $\gamma $,
are equal to zero for $W<W_{c}$ and equal to unity for  $W>W_{c}$.

Fig.~\ref{fig2} shows the disorder dependences of $\alpha$ and $\gamma$.
The curves for different sizes cross
at one point, $W_{c}$, which corresponds to the disorder-induced MIT.
Near the critical point, as long as $|W-W_{c}| \lesssim 2$,
$\alpha(W)$ and $\gamma(W)$ depend stronger on $W$ for larger cubes.
A similar behavior of $\gamma$ was observed
earlier~\cite{Shklovskii93} for much smaller cubes than ours.
A different variable,
$\int_{s_{2}}^{\infty}\,I(s)\,ds+s_{2}-1$,
was studied in~\cite{Hofstetter94}, which corresponds to using $P(s)$
in the range of large spacings, as $\gamma $ does.
A comparison of the two plots in Fig.~\ref{fig2} shows
that near the MIT the scaling variables $\alpha$ and
$\gamma$ behave almost identically.
The critical disorder and the scaling properties of the integrated
level spacing distribution do not depend on the choice of $s_{o}$.
We have checked this $s_{o}$-independence by repeating
the calculations for $s_{o}$=0.1 and 4.
However, the results for $s_1$ yields the highest accuracy.
Using the data for $\alpha $ one can determine very accurately
the critical value of the disorder, $W_{c}=16.35\pm 0.15$.

In the following we discuss the critical behavior of $\alpha $.
It characterizes the slope of the linearly growing part of $P(s)$.
Assuming the validity of the one-parameter scaling
$\alpha(W,L)=f(L/\xi(W))$,
one can replot the  data of Fig.~\ref{fig2} by rescaling $L$.
We numerically chose values of $\xi$ by shifting
the logarithm of the length separately for each $W$.
An overlap between data for adjacent values of $W$ allows to fit
most of the points onto a common curve with two branches,
one growing  ($W>W_{c}$) and one decaying  ($W<W_{c}$) (Fig.~\ref{fig3}).
Thus we find a function $\xi(W)$ up to an arbitrary factor $\xi_{o}$.
This arbitrariness can be avoided.
The points for the maximum disorder $W=20$ in Fig.~\ref{fig3} except of the
sizes $L=6$ and 8 correspond to the localized regime $\xi \lesssim L$.
Supposing that the statistics of the levels of localized states
is not sensitive to the dimensionality of system,
one can match  $\xi$ for the above disorder
to the relevant values of the localization length
calculated by the TM-method in the quasi-1D case ~\cite{Kramer83}.
The function $\xi (W)$, as found by this procedure,
is plotted in the inset of Fig.~\ref{fig3}.
Our values of $\xi$ coincide
with those found by the TM-method~\cite{Kramer83,KramerM94}.
The agreement is slightly better for the insulating
than for the metallic side.

The next step is to find the critical exponent.
Using the singularity of the correlation length near $W_{c}$
\begin{equation}
\xi(W) = l \left| \frac{W-W_{c}}{W_{c}} \right| ^{-\nu},
\label{Loclen}
\end{equation}
where the scale $l$ is of order of the lattice constant, one can expand
the scaling function $f(L/\xi)$ into a power series
$\alpha(W,L) = \alpha_{c}(L) + A (W-W_{c}) L^{1/\nu}$,
By applying a conventional $\chi^2$-criterion
for the verification of the linearized scaling hypothesis
with several parameters to fit the data in Fig.~\ref{fig2}
we found  $\nu=1.45\pm 0.08$.
It is well consistent with the critical exponent $\nu\simeq 1.53$
obtained recently by the TM-method~\cite{MacKinnon94}.
A slightly smaller value of $\nu\simeq 1.34$
for the large-$s$ region of $P(s)$ has been computed
in~\cite{Hofstetter94} by using the level-statistics method similar to ours.
The dependence $\xi(W)$ from (\ref{Loclen}) with $\nu=1.45$ is shown
in Fig.~\ref{fig3} for comparison.
We note that the critical exponent could be also estimated
by fitting the data for $\xi^{-1}(W)$ to the power-law function
without linearizing of $\alpha(W)$ near $W_{c}$.
However the relative numerical errors become markedly larger
similar as in the earlier approach~\cite{Kramer83}.
The accuracy of $\nu$ can then be improved by considering a narrower interval
of $W$ near the critical point or by using larger systems.

In conclusion, we have studied the integrated level spacing distribution
as a scaling variable in order to detect the critical behavior of the
statistics of the one-electron levels at the
disorder-induced metal-insulator transition.
The Anderson model was used to calculate the spectra of disordered
cubic lattices of sizes varying from $6^{3}$ up to $28^{3}$ sites.
Diagonalization of the Hamiltonian was performed
numerically by means of the Lanczos
algorithm especially optimized for sparse symmetric matrices with a
hierarchic structure.
We calculated the critical distribution $P_{c}(s)$
at the transition and confirmed its size-independence
conjectured in~\cite{Shklovskii93}.
Near the critical point the finite-size scaling properties
of the probability $I(s)$ of neighboring spacings were examined.
For the small-$s$ and the large-$s$ parts of $P(s)$ this probability
exhibits the transition
between the Wigner and the Poisson distributions which corresponds to
the delocalization-localization transition.
Using small spacings the critical disorder
was found, $W_{c}=16.35$, for
the `box distribution' of the site energies.
Finally, we determined the disorder dependence of the correlation length
$\xi(W)$ which is in good agreement with
previous results obtained by
the transfer-matrix method~\cite{KramerM94}.
The present calculation confirms one-parameter scaling by
using a completely independent method, and that the critical exponent
of the purely disorder-induced
metal-insulator transition is $\nu =1.45\pm0.08$.

We thank
B.~I. Shklovskii, T.~Brandes, M.~Schreiber and A.~Mac\,Kinnon for discussions.
Financial support from DAAD during the stay of I.~Kh.~Zh. at the  University
of Hamburg is gratefully acknowledged.
This work was supported by EU via grants SCC-CT90-0020 and
CHRX-CT93-0126.

\newpage

\begin{figure}[t]
\caption[]{Level spacing distributions $P(s)$
for cubes of sizes $L^3=6^3$ and $28^3$
for various values of the disorder $W$ near the transition.
Continuous curves are the Wigner surmise, Eq.~(\ref{Wigner}), and
the Poisson distribution, Eq.~(\ref{Poisson}).}
\label{fig1}
\end{figure}

\begin{figure}[t]
\caption[]{Scaling variables  $\alpha(W)$ and $\gamma(W)$
near $W_{c}$ for different
$L = 6 (\Diamond)$, $8 (+)$, $12 (\Box)$, $16 (\times)$,
$20 (\bigtriangleup)$, $24 (\star)$, $28 (\circ)$.}
\label{fig2}
\end{figure}

\begin{figure}[t]
\caption[]{Scaling variable $\alpha$ as a function of $L/\xi(W)$,
showing one-parameter scaling.
$W = 12 (\Diamond)$, $14 (+)$, $15 (\Box)$, $16 (\times)$,
$16.5 (\bigtriangleup)$, $17 (\star)$, $18 (\circ)$, and $20 (\bullet)$.
Inset: correlation length $\xi$ as function of the disorder $W$.
Present method ($\bullet$),
TM-method ($\Diamond$)~\cite{Kramer83}; (+)~\cite{KramerM94}.
Continuous curve: plot of Eq.~(\ref{Loclen}) with
$W_{c}=16.35$, $\nu=1.45$, and $l=1$.}
\label{fig3}
\end{figure}

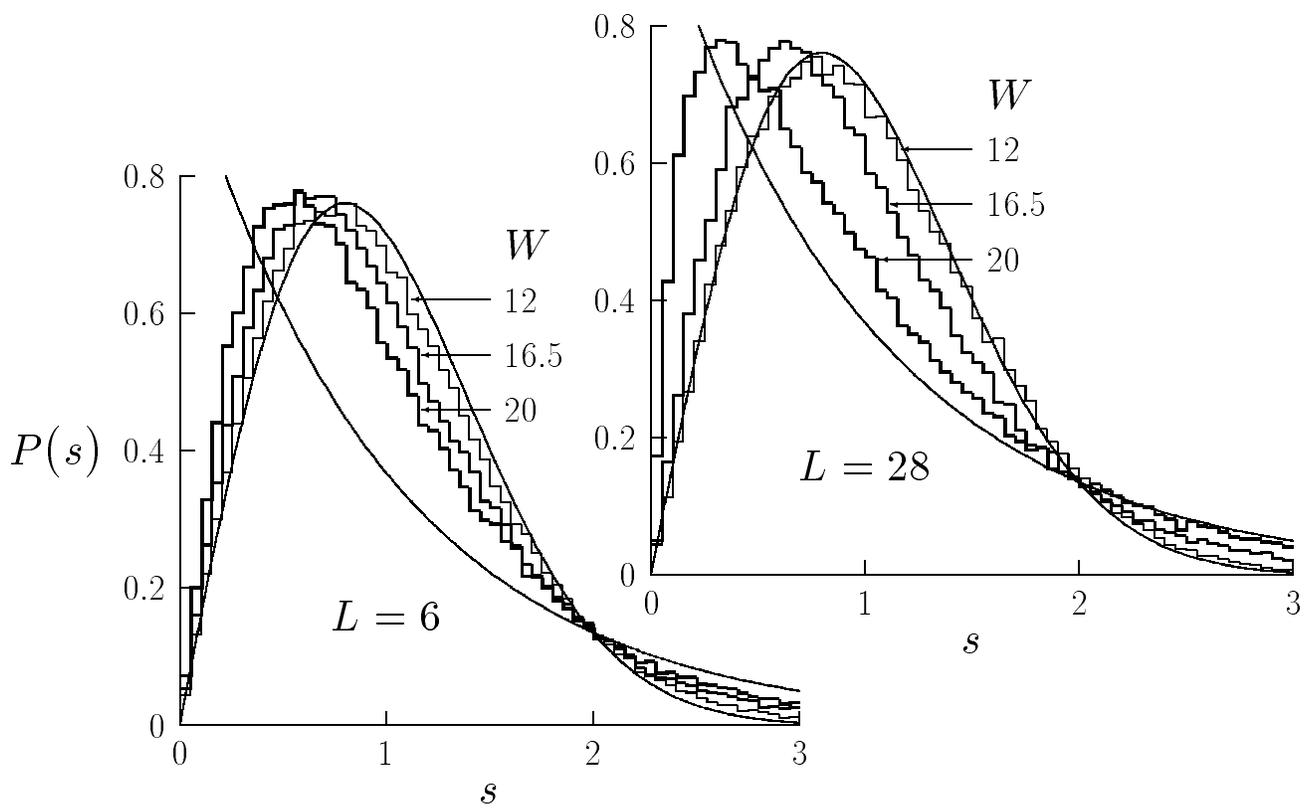
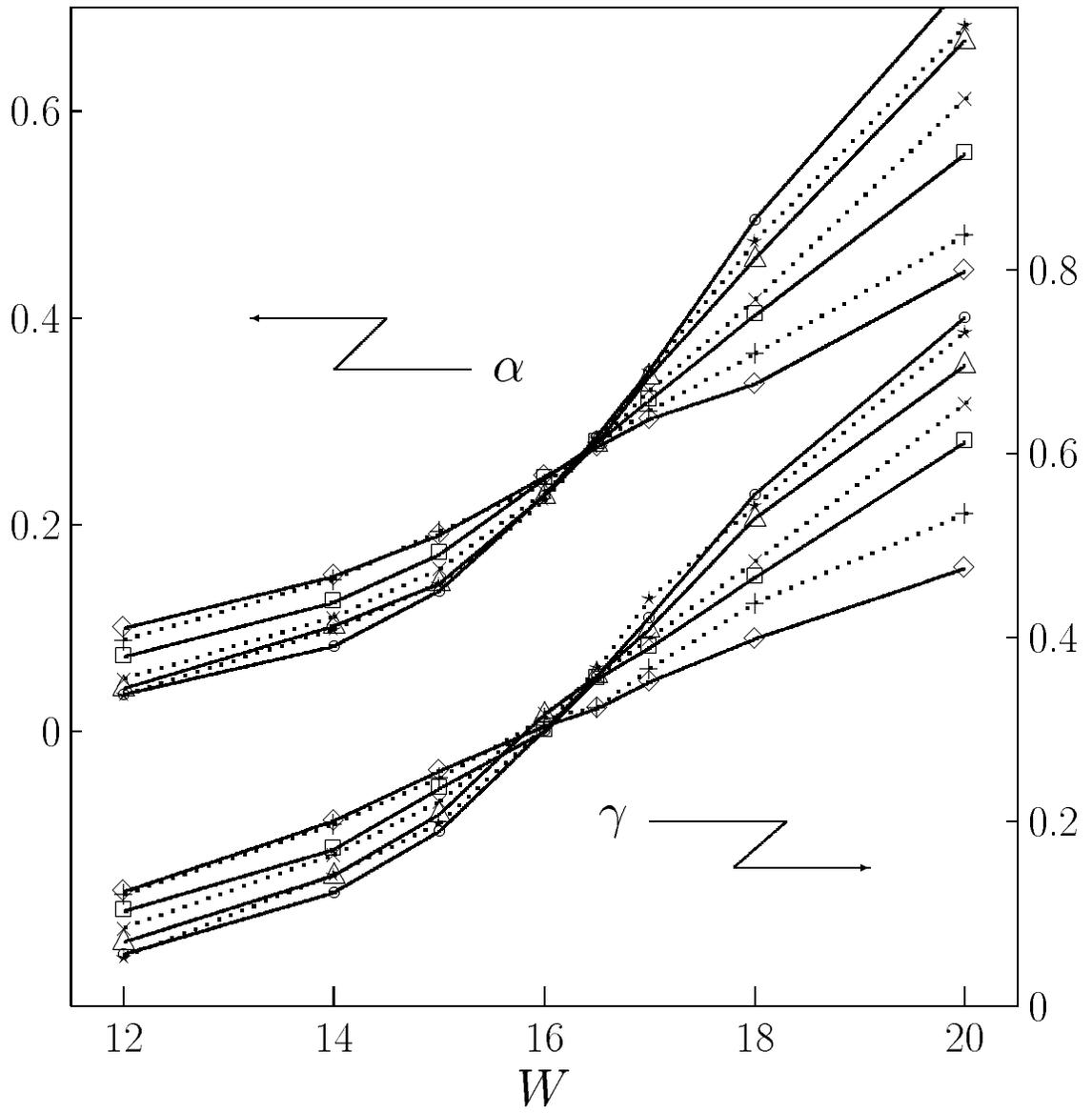
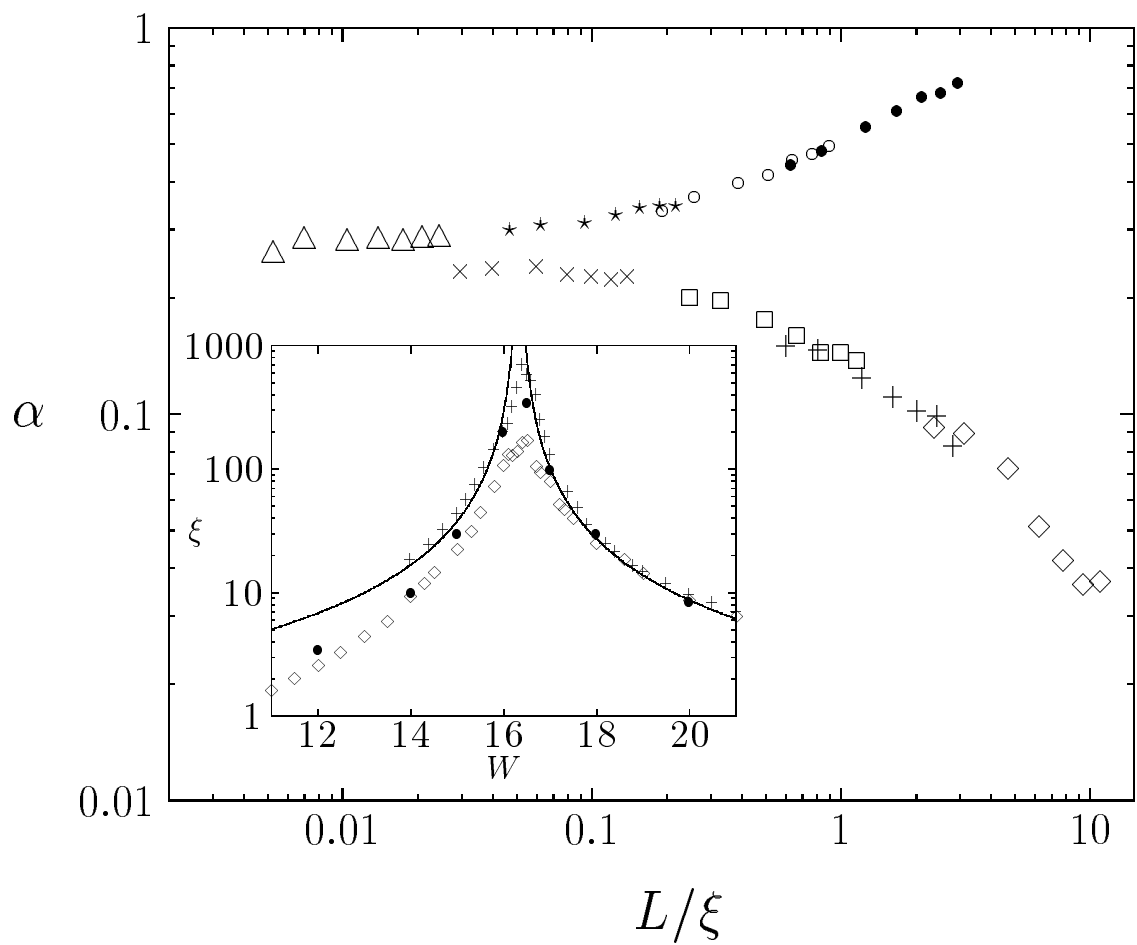
\end{document}